\documentclass[aps,pre,reprint,groupedaddress,floatfix,longbibliography]{revtex4-1}

\usepackage{amsmath}
\usepackage{graphicx}

\begin{document}

\title{Enhanced non-resonant light transmission through subwavelength slits in metal}

\author{Anders Pors}
\email[]{alp@iti.sdu.dk}
\affiliation{SDU Nano Optics, University of Southern Denmark, Campusvej 55, DK-5230 Odense M, Denmark}

\author{Khachatur V. Nerkararyan}
\affiliation{Department of Physics, Yerevan State University, 375049 Yerevan, Armenia}

\author{Khachik Sahakyan}
\affiliation{Department of Microwave Physics and Telecommunications, Yerevan State University, 375049 Yerevan, Armenia}

\author{Sergey I. Bozhevolnyi}
\affiliation{SDU Nano Optics, University of Southern Denmark, Campusvej 55, DK-5230 Odense M, Denmark}







\begin{abstract}
We analytically describe light transmission through a single subwavelength slit in a thin perfect electric conductor screen for the incident polarization being perpendicular to the slit, and derive simple, yet accurate, expressions for the average electric field in the slit and the transmission efficiency. The analytic results are consistent with full-wave numerical calculations, and demonstrate that slits of widths $\sim 100$\,nm in real metals may feature non-resonant (i.e., broadband) field enhancements of $\sim 100$ and transmission efficiency of $\sim 10$ at infrared or terahertz frequencies, with the associated metasurface-like array of slits becoming transparent to the incident light.
\end{abstract}



\maketitle

General perception of light transmission through deeply subwavelength apertures in opaque metal screens was first considerably changed back in 1998, when the phenomenon of extraordinary optical transmission (EOT) was introduced \cite{ebbesen_1998}. Previously, it was generally accepted that subwavelength apertures, similar to small particles, only weakly interact with the incident light, exhibiting progressively lower transmission for the apertures becoming smaller than the light wavelength. In the EOT case, which typically occurs at visible and near-infrared wavelengths, periodic arrays of apertures (or isolated apertures surrounded by periodic structures \cite{thio_2001}) facilitate constructive excitation of surface plasmon polaritons that transpires to resonant transmission of light, with apertures demonstrating a normalized-to-area transmission considerably larger than one. 

We note that the work done in relation to EOT is vast, and for a detailed overview of past achievements we refer to a selection of comprehensive review papers \cite{genet_2007,abajo_2007,coe_2008,vidal_2010}. In this Letter, we revisit the case of light transmission through an isolated one-dimensional subwavelength slit in an otherwise opaque metal screen that can be treated as a perfect electric conductor (PEC).  This kind of configuration has been subject to analytical, numerical, and experimental treatments in the past, but we emphasize that in most of those studies extraordinary transmission occurs due to either standing-wave resonances in thick metal screens \cite{takakura_2001,yang_2002,gordon_2006} or shape resonances in wide slits \cite{vidal_2003,lee_2007,lee_2009}, with the former type of resonance also being utilized in realizing perfect endoscopes in arrayed counterparts \cite{jung_2009}. Here, we focus on subwavelength-thin metal screens for which no resonances exist, and we present accurate, yet simple, analytical formulas for key parameters, like the average field enhancement in the slit and the transmission efficiency. The analytical description is consistent with numerical calculations and related theoretical \cite{novitsky_2011} and experimental work \cite{seo_2009}, hence establishing insight into the process of passing light through slit apertures. 

\begin{figure}[tb]
	\centering
		\includegraphics[width=8.1cm]{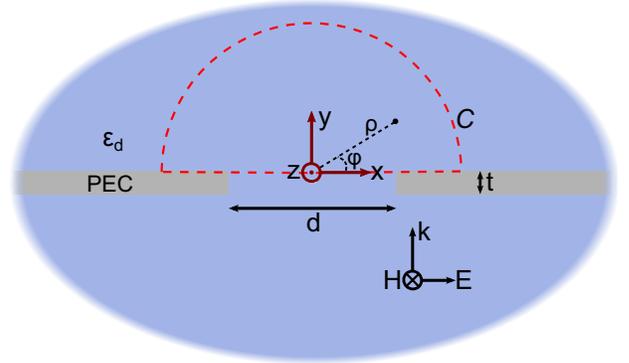}
	\caption{Sketch of slit in a PEC screen defined by the width $d$ and thickness $t$. The surrounding medium is characterized by the relative permittivity $\varepsilon_d$, and the incident TM-polarized plane wave propagates along the $y$-direction.}
	\label{fig:1}
\end{figure}

\begin{figure}[tb]
	\centering
		\includegraphics[width=8.8cm]{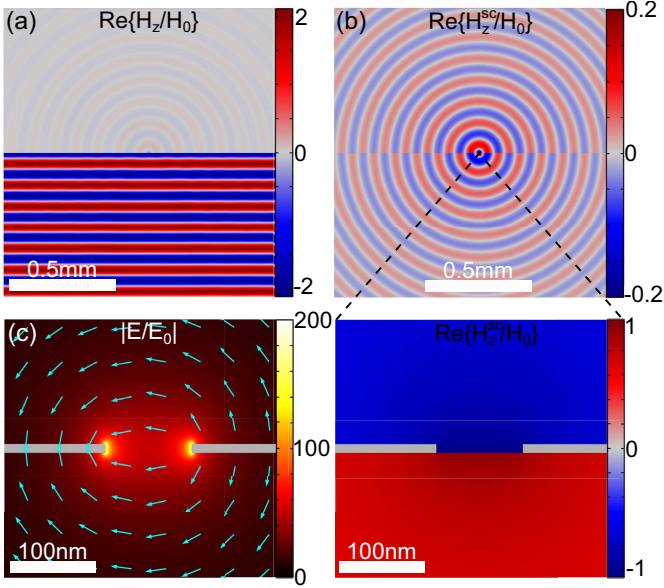}
	\caption{Color maps of the (a) total and (b) scattered magnetic field in the vicinity of the slit for the parameters $t=10$\,nm, $d=100$\,nm, and $\lambda=100$\,$\mu$m. (c) Color map of the total electric field near the slit, together with arrow plot showing the direction of the field. Note that all plots are normalized to the value of the incident wave, while in (b,c) the scale bars are chosen to emphasize to the distribution of the fields.}
	\label{fig:2}
\end{figure}

The considered configuration represents a one-dimensional slit of subwavelength width $d$ in a PEC film of thickness $t$ (Fig. \ref{fig:1}). The incident field is chosen to be a transverse magnetic (TM) polarized plane wave (i.e., $H_z,E_x,E_y\neq 0$) propagating normal to the film surface, with electric and magnetic field components defined by
\begin{equation}
E^{\mathrm{in}}_x(y)=E_0e^{iky} \quad, \quad H^{\mathrm{in}}_z(y)=-E_0 Z^{-1} e^{iky}.
\label{eq:in}
\end{equation}
Here, we implicitly assume harmonic-time dependence $\exp(-i\omega t)$, $k=2\pi/\lambda$ is the wave number in the surrounding medium of permittivity $\varepsilon_d$, $\lambda=\lambda_0/\sqrt{\varepsilon_d}$, $\lambda_0$ is the free-space wavelength, $E_0$ is the amplitude of the incident electric field, and $Z$ is the wave impedance.
%
In order to proceed, we assume the relation $t\ll d\ll\lambda$, meaning that in the analytical description of light transmission through the slit it is reasonable to neglect the film thickness (i.e., $t=0$\,nm) and for $\rho>\lambda$, where $\rho=\sqrt{x^2+y^2}$, to treat the slit as a point scatterer. For a homogeneous, isotropic domain for which an arbitrary field solution can be represented by a series of cylindrical wave functions \cite{stratton}, the latter assumption implies the following functional form of the the scattered field components in each half-space
\begin{equation}
E^{\mathrm{sc}}_\varphi(\rho)=BH_1^{(1)}(k\rho) \quad , \quad H^{\mathrm{sc}}_z(\rho)=-iB Z^{-1} H_0^{(1)}(k\rho) \quad (\rho\gg d),
\label{eq:sc}
\end{equation}
where $B$ is the amplitude, $H_n^{(1)}$ is the Hankel function of first kind and order $n$, and $E_\phi=-E_x\sin\varphi+E_y\cos\varphi$, where the angle $\varphi$ is measured in the counter-clock direction from the $x$-axis (see Fig. \ref{fig:1}). It is worth noting that the cylindrical symmetry of the scattered light (i.e., independent of $\varphi$) is a consequence of the point scatterer assumption. Full-wave numerical simulations, performed using the commercially available finite element software Comsol Multiphysics, however verify the approximation, as seen in Fig. \ref{fig:2}(b) for the scattered magnetic field. Moreover, it is evident that the only difference between the two half-space solutions is a change of sign.   
%

We would now like to relate the amplitude of the scattered light $B$ to the average electric field in the slit, denoted $A$. In doing so, we approximate the scattered electromagnetic near-field close to the slit with Eq. \eqref{eq:sc} for small arguments of the Hankel functions, i.e. 
\begin{equation}
E^{\mathrm{sc}}_\varphi(\rho)\simeq \frac{-2iB}{\pi k\rho} \quad , \quad H^{\mathrm{sc}}_z(\rho)\simeq \frac{-iB}{Z}\left[1+\frac{2i}{\pi}\ln(k\rho)\right] \quad (\lambda \gg\rho\gg d).
\label{eq:scnear}
\end{equation}
It should be noted that the expansion of $H_0^{(1)}$ for $k\rho\rightarrow 0$ typically only contains the dominant imaginary logarithm term. However, we would like to point out that the inclusion of a constant real part in $H_0^{(1)}$ (1-term in square brackets) is helpful for an accurate description of the phase of the scattered near-field. 
For a series expansion up to first order in $k\rho$, the imaginary part of $H_0^{(1)}(k\rho\rightarrow 0)$ actually also features a constant of $\simeq-0.074$, which we have neglected due to its small value.   

In the next step of the analytical treatment, we consider the electric near-field in the electrostatic limit. Since this limit implies $\nabla \times \mathbf{E}^{\mathrm{sc}}=0$, it directly follows from Stokes theorem that 
$$ \oint_C\mathbf{E}^{\mathrm{sc}}\cdot \mathrm{d}\mathbf{l}=0,$$
where the path $C$ corresponds to a semi-circle in the upper half-space for which $\lambda\gg\rho\gg d$ (see Fig. \ref{fig:1}). The line integral can be evaluated, yielding 
\begin{equation}
E_{\varphi}^{\mathrm{sc}}=-\frac{dA}{\pi\rho} \quad (\rho\ll \lambda),
\label{eq:scnear2}
\end{equation} 
where
\begin{equation}
A=\frac{1}{d}\int_{-d/2}^{d/2}E_x^{\mathrm{sc}}(x,y=0)\,\mathrm{d}x.
\label{eq:A}
\end{equation}
We note that Eqs. (\ref{eq:scnear}) and (\ref{eq:scnear2}) allow us to relate the amplitude of the scattered field to the average electric field in the slit,
\begin{equation}
B=\frac{-i\pi d}{\lambda}A.
\label{eq:B}
\end{equation}
With the above equation, the last task is concerned with relating the amplitude of the incident field $E_0$ to the average field in the slit $A$. We note that the condition $d\ll\lambda$ entails a scattered electromagnetic field away from the slit that is weak compared to the incident field. This fact is illustrated in Fig \ref{fig:2}(a), thus verifying that the dominant field in the lower half-space is the standing wave pattern arising from the interference of incident and specular reflected waves, i.e., 
\begin{equation}
E_x^{(-)}(y)=2iE_0\sin(ky) \quad,\quad H_z^{(-)}=-2E_0 Z^{-1}\cos(ky).
\label{eq:lower}
\end{equation}
It is evident that the electric field is zero at the material interface ($y=0$), hereby implying that continuity of $E_x$ through the slit transpires to continuity in $E_x^\mathrm{sc}$. This has the consequence that $E_x^\mathrm{sc}(x,y)=E_x^\mathrm{sc}(x,-y)$, as seen in the arrow plot of Fig. \ref{fig:2}(c). Moreover, as $E_x^\mathrm{sc}$ is the all-dominant electric field component in the slit and decays along the $y$-axis away from the slit, the scattered magnetic field $H_z^\mathrm{sc}$, which is proportional to $\partial E_x^{sc}/\partial y$, must satisfy the equality $H_z^\mathrm{sc}(x,y)=-H_z^\mathrm{sc}(x,-y)$ [see Fig. \ref{fig:2}(b)]. The continuity of the total magnetic field at the slit is represented by the equation $H_z^{\mathrm{sc}}= H_z^{(-)}- H_z^{\mathrm{sc}}$, which under the assumption that the magnetic field is constant in the slit [Fig. \ref{fig:2}(b)] entails $H_z^{\mathrm{sc}}=-E_0/Z$ in the slit.

In the final step of relating $E_0$ to $A$, we invoke the field equivalence principle for apertures in PEC screens. Here, it is known that scattering from the aperture is equivalent to radiation from a \emph{magnetic} current $\mathbf{J}_m=-2\hat{\mathbf{n}}\times\mathbf{E}_a$, where $\hat{\mathbf{n}}$ is the surface normal and $\mathbf{E}_a$ is the electric aperture field \cite{schelkunoff_1939}. For the subwavelength slit considered in this Letter, the magnetic current can be approximated by a point source with $\mathbf{J}_m=2Ad\hat{\mathbf{z}}$. Concerning the electromagnetic fields in Eqs. (\ref{eq:sc}) and (\ref{eq:scnear}), we emphasize that the equivalence is only valid for $\rho\gg d$. Nevertheless, in order for the fields to have the same amplitude and phase in the valid regime, the following equality must hold
$$
\frac{2}{d}\int_0^{d/2}H_z^{\mathrm{sc}}\,\mathrm{d}\rho=-\frac{E_0}{Z},
$$
which relates the average magnetic field of the point source within $\rho\leq d/2$ to the same quantity in the slit. By utilizing the near-field expression of $H_z^{\mathrm{sc}}$ in Eq. \eqref{eq:scnear}, performing the integration, and using the relation in Eq. \eqref{eq:B}, we arrive at the result
\begin{equation}
A=\frac{\lambda}{d}\frac{i}{i\pi+2\left[\ln\left(\frac{\lambda}{\pi d}\right)+1\right]}E_0.
\label{eq:A2}
\end{equation}   
As expected, the utilization of dispersion-free materials and a negligible film thickness implies an average field enhancement in the slit ($A/E_0$) that only depends on the ratio $\lambda/d$.

Having derived relations between the different amplitude coefficients, we can calculate the power transmitted through the subwavelength slit and into the far-field ($\rho\gg\lambda$). In this limit, we may use the asymptotic formulas for Hankel functions of large arguments, which leads to the scattered magnetic field  
\begin{equation}
 H^{\mathrm{sc}}_z(\rho)\simeq \frac{-iB}{Z}\sqrt{\frac{2}{\pi k\rho}}e^{i(k\rho-\pi/4)} \quad (\rho\gg \lambda),
\label{eq:scfar}
\end{equation}
with the associated electric field given by $E_\varphi=i Z H^{\mathrm{sc}}_ze^{-i\pi/2}$. As such, the Poynting vector of the scattered field only features a non-zero $\rho$-component that can be written as $S_\rho=\tfrac{1}{2}Z|H_z^{\mathrm{sc}}|^2$, thus resulting in total transmitted power of
\begin{equation}
P_\mathrm{tr}=\int_0^\pi S_\rho \rho \mathrm{d}\varphi=\frac{\lambda}{2Z}\frac{1}{\pi+\frac{4}{\pi}\left[\ln\left(\frac{\lambda}{\pi d}\right)+1\right]^2}E_0^2.
\label{eq:Ptr}
\end{equation} 
As an interesting figure of merit, it is instructive to normalize the transmitted power to the power incident on the slit, which corresponds to evaluating the efficiency of which the slit transmits light compared to the geometrical size. The quantity is given by 
\begin{equation}
\eta_\mathrm{tr}=\frac{\lambda}{d}\frac{1}{\pi+\frac{4}{\pi}\left[\ln\left(\frac{\lambda}{\pi d}\right)+1\right]^2}.
\label{eq:sigmatr}
\end{equation} 
It is worth noting that since $\lambda/d$ is a faster increasing function than $\ln\left(\lambda/(\pi d)\right)$ for $\lambda>d$, it follows that the efficiency of transmitting light and the average field enhancement (for a fixed slit width) monotonically increases with increasing wavelength. The close-to-linear dependence of the field enhancement on the wavelength for $\lambda\gg d$ is in accordance with experimental results conducted in the terahertz regime \cite{seo_2009}.

\begin{figure}[tb]
	\centering
		\includegraphics[width=8.3cm]{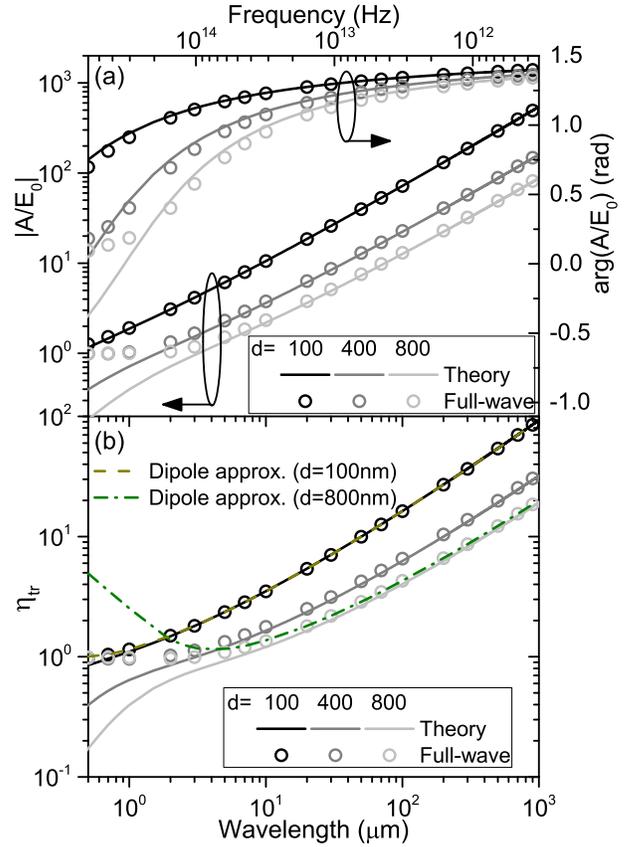}
	\caption{Analytical (lines) and numerical (circles) calculations of the (a) average electric field in the slit and (b) transmission efficiency as a function of wavelength for three values of $d$, keeping $t=10$\,nm in all cases. The dashed lines in (b) correspond to evaluation of $\eta_\mathrm{tr}$ for $d=100$\,nm and $d=800$\,nm using the average electric field from the numerical calculations.}
	\label{fig:3}
\end{figure}

In the remaining part of this work we assess the applicability of the above analytical treatment of light transmission through subwavelength slits in thin metal screens by comparing Eqs. (\ref{eq:A2}) and (\ref{eq:sigmatr}) with full-wave numerical simulations. As an illustrative example, we consider the amplitude and phase of the average electric field at the center line of the slit for the slit widths $d=100$, 400, and 800\,nm in the wavelength range $0.5-1000$\,$\mu$m, keeping the screen thickness fixed at $t=10$\,nm [Fig. \ref{fig:3}(a)]. 
%
Taking into account the approximations involved in the derivation of Eq. \eqref{eq:A2}, it is remarkable how it almost perfectly describes both the amplitude and phase of the average electric field for $\lambda/d>10$ and with only little discrepancy down to $\lambda/d\sim5$. It should be noted that the wavelength-dependent phase response owes to the inclusion of the constant real part of the series expansion of $H_0^{(1)}$ for small arguments [see Eq. \eqref{eq:scnear}]. From the point of view of applications in the infrared (or larger wavelength) regime, the fact that field in a slit (that can be produced with standard nanofabrication techniques) can significantly be enhanced with a proper choice of $\lambda/d$ by making use of Eq. \eqref{eq:A2} is also rather encouraging. The associated transmission efficiencies are found [Fig. \ref{fig:3}(b)], in agreement with numerical calculations for $\lambda/d>10$, to exhibit the remarkable phenomenon of significantly enhanced transmission ($\eta_{\mathrm{tr}}\gg 1$) exhibiting a counterintuitive trend of the efficiency becoming larger when increasing the ratio $\lambda/d$. As a way of benchmarking the equivalence between scattering from the slit and a point source, the dashed lines in Fig. \ref{fig:3}(b) show the transmission efficiency for $d=100$\,nm and $d=800$\,nm calculated using the average field in the slit from numerical calculations. As expected, the point scatterer assumption breaks down for $\lambda/d\leq 5$, as particularly evident for $d=800$\,nm.

\begin{figure}[tb]
	\centering
		\includegraphics[width=7.6cm]{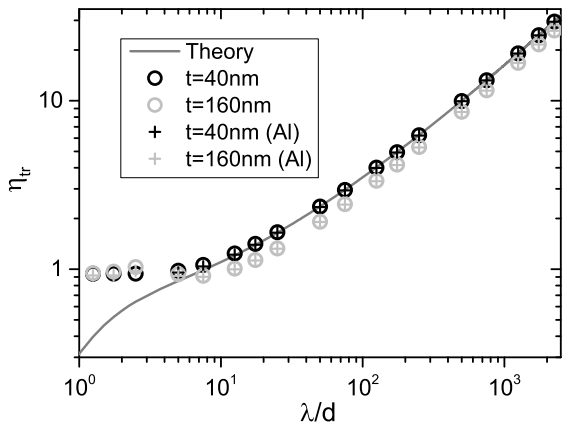}
\caption{Analytical (line) and numerical (circles) calculations of the transmission efficiency as a function of wavelength for $t=40$\,nm and 160\,nm, keeping $d=400$\,nm in both cases. The + markers correspond to the slit transmission efficiency in an aluminum screen.}	
	\label{fig:4}
\end{figure}

In the above analytical derivation, it was assumed that the screen thickness was much smaller than the slit width (i.e., $t\ll d$). In order to gauge the sensitivity of the transmission on $t$, Fig. \ref{fig:4} displays the numerically obtained transmission efficiency when $d/t=10$ and $2.5$ and compared with the (thickness-independent) analytical result. 
%
It is evident that when the condition $t\ll d$ is not strictly satisfied, the analytical result slightly overestimates the transmission efficiency, though the wavelength-dependence is still well-captured for $\lambda \gg d$. Moreover, Fig. \ref{fig:4} demonstrates that the transmission efficiency of a slit in an aluminum screen, described by a Drude permittivity with plasma and collision frequency $\hbar\omega_p=14.75$\,eV and $\hbar\gamma=81.8$\,meV \cite{ordal_1985}, respectively, is practically equal to the PEC case. For this reason, we expect the presented analytical results to be applicable for a wide range of metals from the mid-infrared regime and lower frequencies. 

\begin{figure}[tb]
	\centering
		\includegraphics[width=7.6cm]{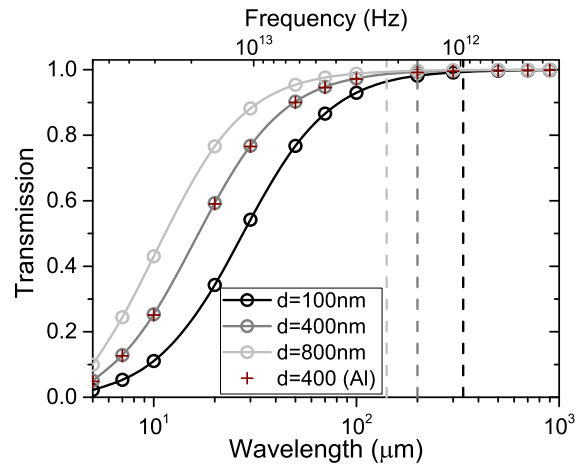}
\caption{Numerical calculation of transmission through slit array as a function of wavelength when $t=d/10$, $d=100-800$\,nm, and $\Lambda=4$\,$\mu$m. The + markers correspond to the transmission in an aluminum screen for $d=400$\,nm. Vertical dashed lines indicate the theoretically predicted onset of perfect transmission.}	
	\label{fig:5}
\end{figure}

As a final application of the presented theory, we consider an array of slits with period $\Lambda$ and satisfying the inequality $t\ll d \ll \Lambda \ll \lambda$. The configuration may be considered as a metasurface, while the condition $d \ll \Lambda$ ensures that the interaction between neighboring slits is weak, thus making the response of the metasurface related to the properties of the individual slits. Within this line of reasoning, the array of slits should display (non-resonant) perfect transmission when 
\begin{equation}
\eta_{tr}d \geq \Lambda.
\label{eq:t1}
\end{equation} 
Our full-wave numerical simulations of the radiation transmission through an array of slits in PEC and aluminum screens demonstrate that the array transmission rapidly increases as a function of wavelength (Fig. \ref{fig:5}). 
Interestingly, these metasurface-like configurations do reach non-resonant perfect transmission at the wavelengths that are very close to the values predicted by $\eta_{tr}d = \Lambda$. 

In conclusion, we have derived simple, yet accurate, formulas for the average electric field and transmission efficiency of light passing through a subwavelength slit in a thin PEC screen. The formulae obtained are applicable for practically all metals from the mid-infrared regime and at lower frequencies. For example, slits that can be fabricated using standard nanofabrication techniques may feature field enhancement of $\sim 10^2$ and transmission efficiency of $\sim 10$ in the infrared or terahertz regime, while the associated array of slits would display non-resonant perfect transmission. We believe that our findings have important implications to various fundamental and technological applications ranging from near-field THz microscopy with subwavelengths slits to surface-enhanced linear and nonlinear spectroscopy and sensing \cite{tonouchi_2007}.

%
%
%
%
%
%

\section*{Funding Information}

\textbf{Funding.} European Research Council (341054); Danish Council for Independent Research (1335-00104A); University of Southern Denmark (SDU 2020 funding).


%
%
%
%
%

\end{document}